\documentclass[prb,aps,preprint]{revtex4-2}
\usepackage{epstopdf}
\usepackage{graphicx}  
\usepackage{dcolumn}   
\usepackage{bm}        
\usepackage{amssymb}
\usepackage[english]{babel}
\usepackage{lineno}

\begin{document}

\title{Thermal conductivity of $iron$ and $nickel$ during melting: Implication to Planetary liquid outer core}

\author{Pinku Saha$^{1,2}$}

\author{Goutam Dev Mukherjee$^1$}
\email [Corresponding Author:]{goutamdev@iiserkol.ac.in}

\affiliation {$^1$National Centre for High Pressure Studies, Department of Physical Sciences, Indian Institute of Science Education and Research Kolkata, Mohanpur Campus, Mohanpur – 741246, Nadia, West Bengal, India.}

\affiliation{$^2$Chemistry and Physics of Materials Unit, Jawaharlal Nehru Centre for Advanced Scientific Research, Jakkur P.O., Bangalore 560064, India}


\begin{abstract}
We report the measurements of the thermal conductivity ($\kappa$) of $iron$ ($Fe$) and $nickel$ ($Ni$) at high pressures and high temperatures. $\kappa$ values are estimated from the temperature measurements across the sample surface in a laser heated diamond anvil cell (LHDAC) and using the COMSOL software. Near-isothermal $\kappa$'s are observed to increase with pressure in both the metals due to the increase of density of the pressed metals. In both metals $\kappa$'s are observed to follow a sharp fall during melting at different pressure points and are consistence with the other multi-anvil measurements. Constant values of $\kappa$ in these metals during melting at different pressures reveal the loss of long range order, which creates independent movement of atomic metals. The melting temperature measured in these metals from the sudden drop of $\kappa$-values are in a good agreement with the other melting measurements in LHDAC. The results obtained in this study is expected to provide an insight to the studies on the planets Mercury and Mars and their interior.

\noindent {\bf Keywords:}{Laser heated diamond anvil cell, COMSOL, Thermal conductivity, High pressure effects, Constant thermal conductivity, Geodynamo}

\end{abstract}

\maketitle

\section{Introduction}

Seismological and geophysical studies revealed that the elements having high binding energy like $Fe$ and $Ni$ in their pure phase or in the alloy form with light elements (C, S, Si, and O etc.) are present in the planetary core\cite{birch52,dziewonski81,stevenson81,mao90,wood93,poirier94,dubrovinsky00,li02,li03,nakajima09,sata10,mookherjee11,nakajima11,chen12,chen14,prescher15,liu16,saha20}. These materials are predicted to be present in liquid state at the outer core of the planets. The knowledge of the transport properties of $Fe$ and $Ni$ at high pressure (HP) and high temperature (HT) conditions are essential for better understanding the generation of the magnetic field and the heat loss from Earth's core. Therefore, the measurement of the thermal conductivity ($\kappa$) of these materials at extreme conditions of pressure and temperature are very important for the understanding the dynamics of the planetary interior. The value of the $\kappa$ determines the age of the inner core (low value implies older core). Stacey {\it et al.} predicted a lower value of $\kappa$$\sim$ 28-29 $W m^{-1} K^{-1}$ from the theoretical calculations on $Fe-Ni-Si$ alloy\cite{stacey01,stacey07} by extrapolating near ambient conditions data. Later the first principles calculations using density functional theory reported the value of $\kappa$ to be very high $\sim$ 160-200 $W m^{-1} K^{-1}$\cite{sha11,alfe12,deKoker12,pozzo12,pozzo14,davies15} at the Earth's core conditions.In their studies mostly they considered the electron and phonon contribution terms independently for the anharmonicity effect and the change in the Fermi surface, respectively.
Recently Xu {\it et al.}\cite{xu18} computed $\kappa$ to be 77$\pm$10 $W m^{-1} K^{-1}$ at the Earth's outer core conditions considering both the electron-phonon ({\it e-ph}) and electron-electron ({\it e-e}) scattering contributions to electrical and thermal conductivity in solid $hcp-iron$. The thermal conductivity in the outer-inner core boundary on C, O, Si and S doped Fe + 10\% Ni at a rate 30\% ternary systems were calculated in the range of 105 – 140 $W m^{-1} K^{-1}$ very recently by Zidane et al\cite{zidane20}. The above study did not discuss about the Bloch-Gruneisen law in combination with Mathiessen’s rule to describe the phononic contribution, however they claimed that upon consideration of the phononic contribution, the thermal conductivity will decrease.

The experimental determination of the thermal conductivity at extreme conditions of pressure and temperature are very challenging and rare\cite{konopkova11,gomi13,ohta16,konopkova16,sahagf20}. Thermal conductivity of $Fe$ was determined directly from the measurements of the temperature and heat propagation\cite{konopkova11,konopkova16,sahagf20} in a laser heated diamond anvil cell (LHDAC). There are several reports of $\kappa$, estimated indirectly, from the high pressure and high temperature electrical resistivity measurements using multi anvil cell (MAC)\cite{deng13,secco17,silber17,silber18,ezenwa19} and diamond anvil cell (DAC)\cite{gomi13,ohta16,basu20}. The $\kappa$ was estimated from the resistivity ($\rho$) data employing Wiedemann-Franz-Lorenz law. From resistivity measurement, Gomi {\it et al.} predicted $\kappa$ for $Fe$ to be higher than 90 $W m^{-1} K^{-1}$ at Earth's outer core conditions from DAC experiments combined with theoretical calculations\cite{gomi13}. Later, from the resistivity measurements in LHDAC, Ohta {\it et al.} estimated a high value of $\kappa$ ($\sim$ 226 $^{+72}_{-31}$ $W m^{-1} K^{-1}$) of $Fe$ at core mantle boundary conditions based on resistivity saturation of $hcp-Fe$. These diamond cell experiments are well supported by the theoretical calculations\cite{sha11,alfe12,deKoker12,pozzo12,pozzo14,davies15}. In contrast to the above studies all the MAC experiments reported $\kappa$ values in the range 39-70 $W m^{-1} K^{-1}$ for solid and liquid $Fe$ and $Ni$ up to 15 GPa\cite{deng13,silber17,silber18,secco17,ezenwa19}. In these MAC experiments a sudden drop ($\sim$ 20-40\%) of $\kappa$ at the melting temperatures were observed and molten metals showed a constant resistivity over their experimental pressure scale. Studies carried out using LHDAC with pulsed laser, Konopkova {\it et al.} reported a low value of thermal conductivity of $Fe$ about 35 $W m^{-1} K^{-1}$ at 48 GPa and 2000 K, and in the range 18 - 44 $W m^{-1} K^{-1}$ from 48 GPa to  130 GPa and temperature range 2000 -3000 K, respectively \cite{konopkova16}. In their previous work they estimated a value around 32 $\pm$ 7 $W m^{-1} K^{-1}$  at 78 GPa and 2000 K using continuous wave (CW) high power infrared laser, and finite-element numerical simulations\cite{konopkova11}. Very recently, a direct measurement reported $\kappa$ to be 70-80 $W m^{-1} K^{-1}$ (with an uncertainty of 40\%) which remains constant in the $hcp$ phase of $Fe$\cite{sahagf20}. In this study the temperature gradient across the sample surface was measured during heating the sample with a continuous wave infrared laser followed by computing the temperature gradient profile using COMSOL software under the steady state heat flow condition. The same study predicted that the effect of high temperature combined with melting of the sample will reduce $\kappa$ value to 40 $\pm$ 16 $W m^{-1} K^{-1}$ at the outer core conditions of Earth. The direct measurements of $\kappa$ during the melting of any Earth's core materials have not been reported so far.       

In the present work, we have carried out the measurements of thermal conductivity of $Fe$ and $Ni$ at high pressures using a single sided LHDAC facility and COMSOL software. The steady state heat flow condition is assumed in the COMSOL software and heat absorbed by the metal foil is calculated using thermodynamical equation. Temperature dependent thermal conductivity of compressed $Fe$and $Ni$ at various pressure points are measured and compared with the literature values to see any changes in their values during melting. The pressure dependence melting temperatures estimated from the observation of the $\kappa$ value anomalies, and  are compared with the literature.

\section {Experimental and Method}
Thermal conductivity measurements at high-pressures are carried out by measuring temperatures across the sample in LHDAC and simulating the temperature gradient profile using COMSOL software\cite{comsol1}. The LHDAC consists of a plate type DAC (Almax-Boehler design), and a diode-pumped Ytterbium fiber optic laser (YLR100-SM-AC-Y11) with central emission wavelength, $\lambda$ = 1.070 $\mu$m (maximum power 100 W). In this study diamond anvils of culet flat 300 $\mu m$ are used. T301 stainless steel gasket of initial thickness 225 $\mu$m is preindented to a thickness of 50 $\mu$m by compressing inside DAC. For containing the sample, and pressure transmitting medium (PTM) a hole of diameter $\sim$ 110 $\mu$m is drilled at the center of the culet impression  with the help of electric discharge machine. Thin plates of $Fe$, and $Ni$ of approximate thickness of about 15 $\mu$m are made by compacting polycrystalline $Fe$, and $Ni$ powders using a 300 ton hydraulic press operating at a pressure about 1.5 GPa. The same procedure is followed for the preparation of NaCl plate of approximate thickness of about 12 $\mu$m, which is placed on the both sides of the metal plates. The metal plates ($Fe$/$Ni$), and the NaCl discs are kept at a temperature of 393 K in an electric oven for six hours to remove any trace of moisture. NaCl plates act as both PTM and thermal insulation to the sample from the diamond culet. Thin pieces of NaCl, and $Fe$/$Ni$-plates of desired size (approximate diameter of about 90-110 $\mu$m) are cut to load  in the LHDAC. The sample is sandwiched in between NaCl inside the central hole of the gasket. A few ruby chips (approximate sizes of about 3-4 $\mu$m) are placed at the edge of the metal plates. Pressure inside the LHDAC before and after heating is determined from the shift of the ruby $R_1$ fluorescence lines\cite{mao86}. For reporting the average value of the pressure is taken. Heating is carried out using the diode-pumped Ytterbium fiber optic laser. Heating geometry and procedure are similar to that described by Saha {\it et al.}\cite{sahagf20}. Temperature of the sample surface is measured by spectraradiometry technique by fitting the Planck's radiation function\cite{planck01} in the wavelength range 650-900 nm \cite{boehler90,mukherjee07,sahaijp20}. Temperature of the hotspot can be estimated within an error of about $\pm$15 K in this study and as described in other previous studies\cite{sahagf20,sahaijp20}. However, for determination of actual temperature error in DAC during melting, Saha {\it et al.}\cite{sahaijp20} carried out measurements of melting temperatures of compressed argon within the error of $\pm$25 K. Therefore, we attribute error in our temperature estimation to be within 50 K. 

Temperatures at different positions are measured by translating the 50$\mu$m pin-hole attached to the spectrometer across the magnified image of the sample surface with a resolution of 1 $\mu$m. The sample is heated  either at its center or one of its edge. We wait for about 5 – 10 minutes to have a constant temperature across the hotspot (focused heating laser beam area) is observed. We have carried out time dependent temperature measurements at hotspot and other position of the sample surface at a few pressure points for both metals. Fig.1 shows the time evolution of temperature at the hotspot and at a position $r_2$ away from the hotspot ($\sim$ 70 $\mu$m) at two pressure points for $Ni$.  Temperatures at same position with time are found to remain constant within the error limit and this ensures the steady-state condition. We start with a model that simulates the steady-state temperature distribution in a cylindrical metal plate continuously heated by a heat source. The equation that the heat conduction in system will follow is:

\begin{equation}
\nabla .(-\kappa \nabla T)=f(r)
\end{equation}
\noindent
where, $\kappa$ is thermal conductivity of the material; $\nabla T$ is the temperature gradient across the sample surface. We have used the boundary conditions for calculation of thermal conductivity of $Fe$/$Ni$ metal in COMSOL\cite{comsol1} and $f(r)$ is the heat source distribution given by:

\begin{equation}
f(r) = Q,  \quad r \leq r_{1}; \; T=T_g \; at \; gasket\; boundary 
\end{equation}

\noindent
Where $r_1$ is the radious of the hotspot, and $T_g$ is the temperature at the gasket boundary. The heat energy $Q$ absorbed by the $Fe$/$Ni$ plate is calculated by
\begin{equation}
Q = mC_p(T_{hotspot}-T_{room})\nu
\end{equation}

\noindent
where, $m$ is the mass of the sample contained at the hotspot, $C_p$ is the specific heat capacity of $Fe$/$Ni$ at constant pressure, $(T_{hotspot}-T_{room})$ is the temperature difference between hotspot and room temperature, and $\nu$ is the modulation frequency (50 kHz) of the 1.070 $\mu m$ wavelength laser. Exposure time of collected  spectrum for temperature measurement is about 100 msec, which is much larger than the modulation period. The specific heat $c_p$ is taken to be 450 JKg$^{-1}$$K^{-1}$ \cite{gubbins03,hirose13,sahagf20}, and 420 JKg$^{-1}$$K^{-1}$\cite{nickelc} for $Fe$ and $Ni$, respectively in our experimental pressure range.

Mass of the sample contained at hotspot $m$ in the Eqn. 3 is calculated as
\begin{equation}
m = \pi r_{1}^{2} h \rho
\end{equation}
\noindent
where, $h$ is the thickness of $Fe$/$Ni$ plate before loading  and $\rho$ is the density of the $Fe$/$Ni$ plate. Initial density of the $Fe$/$Ni$ plate is determined from the weight of the cold compressed plate and measuring its dimensions, which agrees within 95\% with that of a foil. All the used parameter values are listed in the Table-I. As the sample is contained inside the gasket and there is a negligible change in diameter of the gasket hole,  mass of the sample at the hotspot is assumed to remain constant. We find 25\% error in the absorbed power by considering errors in temperature measurement, dimensions of the plate (during mass measurement), and measurement of the hotspot diameter. By taking into account of the errors in every step, we find an uncertainty of about 30\% in determination of thermal conductivity values in $Fe$/$Ni$ plate (Table-II).

We have used finite-element software COMSOL Multiphysics\cite{comsol1} to simulate the temperature distribution in the LHDAC. During the temperature measurements the hotspot along with the sample is covered by PTM around all sides and inside the gasket hole. Three dimensional geometry of the sample chamber and an example of temperature distribution across compressed $Ni$ plate are shown in the Fig.2. Simulation of temperature gradient is carried out to match the experimental temperature profile by varying the thermal conductivity of the sample.

\section{Results and Discussion}
In Fig.3, we have shown the temperature gradient across the $Ni$ plate at three different pressure points (8.7, 17.2, and 22 GPa). Temperature measurements are carried out during heating the sample at the center, which is shown in the right top corner inset (b). The central reddish glow is the hotspot with a temperature 1563 K at 17.2 GPa. On both side of the hotspot, temperatures are measured by translating the spectrometer pinhole with an step size 10$\mu$m. A photograph of the pinhole during measurement is shown in the inset (a) of Fig.3. The filled scattered symbols are the measured temperature, while the solid lines are the computed temperature profile. Thermal conductivity, density values of NaCl and all other geometrical values are given in Table-I. To match temperature gradient with experimental data points, we vary $\kappa$ of $Ni$ plate. An excellent match is seen in the figure. Obtained $\kappa$ values at different pressures and temperatures of the hotspot are indicated in the figure.

In the rest of the measurements we have heated the $Fe$/$Ni$ plate at one of their edge. We have measured $\kappa$ values of compressed $Fe$ plate in its $\gamma$-phase, and the temperature of the hotspot is increased slowly beyond melting temperature following the phase diagram\cite{liu75,anderson86} to see the effect of melting on $\kappa$ values. Temperature dependence of $\kappa$ values of $Fe$-plate at different pressures are shown in Fig.4. Our 8.5 GPa data is compared with that of Saha {\it et al.}\cite{sahagf20} at 10 GPa in Fig.4(a). Both data show a decrease with temperature followed by another sharp decrease after a certain temperature. In Fig.4(c), we have compared the estimated $\kappa$ values with that at 1700 K from Saha {\it et al.}, and at 1250 K from Deng {\it et al.}\cite{deng13,sahagf20} at 7 GPa. Data from Deng {\it et al.} are calculated from the resistivity data by using Wiedemann-Franz law ($\kappa = LT/\rho$; where $\kappa$, L, and $\rho$ are thermal conductivity in $W/m.K$, Lorenz number having value 2.44$\times$10$^{-8}$ $W \Omega K^{-2}$, and electrical resistivity in $\Omega -m$, respectively)\cite{deng13}. It is evident from the Fig.4(c) that the $\kappa$ values in this study are consistent with the previous direct measurements\cite{sahagf20} and seem to agree very well with the indirect measurements of Deng {\it et al.}\cite{deng13}. Also, it can be noted from the Fig.4 that the  $\kappa$ values decrease with temperature and are consistent with previous studies\cite{konopkova16,sahagf20}. Interestingly, we observe a sudden drop in the $\kappa$ values at certain temperatures at different pressures as shown in the figure.  Sharp drops at 5, 7, and 8.5 GPa in $\kappa$ values are observed at temperatures $\sim$ 1975 K, 2035 K, and 2098 K, respectively and are attributed to the melting of the sample at the hotspot while the it is still at solid state at the edge opposite to the hotspot\cite{liu75,anderson86}. These sharp drop in the thermal conductivity values with respect to temperature are due to the presence of the  liquid and solid interface at the boundary of the hotspot. Analogous to this study a sudden jump in the electrical resistivity values were also observed during melting of $Fe$ in high pressure electrical measurements\cite{deng13,ohta16,silber18,basu20}. At all the pressure points, we find 25-30\% decrement in the thermal conductivity values of $Fe$-plate during melting at hotspot.

In Fig.5, we have compared temperature dependent $\kappa$-values of $Ni$ at different pressure points with the electronic thermal conductivity reported in an indirect measurements\cite{silber17} (resistivity) at 4, and 9 GPa in 3000 t MA large volume press. In their study resistivity data were converted to $\kappa$ values by using Wiedemann-Franz law and the Sommerfeld value of the Lorenz number. One can see from the figure that at each pressure the k values show sharp decrease by about 30-35\% at certain temperatures. Sharp decrease of $\sim$40\% is observed by Silber {\it et al.}\cite{silber17} during melting of the $Fe$ sample. Similar behaviour  in the $\kappa$-values are also observed in the case of $Fe$ during melting at the hotspot in this study and previous direct measurements\cite{sahagf20}. Hence we attribute this phenomena to the melting of the sample at the hotspot. A larger drop ($\sim$ 5-10\%) for $Ni$ with respect to $Fe$ in this study may be due to the higher density of $Ni$.  Loss of long range ordering in the molten sample as well as the liquid-solid boundary in our case impede the heat conduction in the sample which in turn result in the low value of thermal conductivity.    
 
The relative change in the thermal conductivity value of $Fe$ with respect to that at 5 GPa are shown in Fig.6(a)while the sample is at molten state at the hotspot area. The ratio shows a constant value of unity demonstrating the invariance of $\kappa$ along the melting boundary. Analogous to this observation, Silber {\it et al.}\cite{silber18} found the constant value of electrical resistivity value in the pressure range 5 to 11 GPa in a MAC and Ohta {\it et al.}\cite{ohta16} in a DAC at 26 GPa for the molten $Fe$. In Fig.6(b), we have plotted $\kappa_P$/$\kappa_0$ (where $\kappa_0$ is thermal conductivity of $Ni$ at ambient pressure calculated from resistivity data of Chu and Chi\cite{chuandchi81}) with pressure during melting of $Ni$-plate and compared with the data of of Silber {\it et al.}\cite{silber17}. In our case the ratio is found to have a constant value around 1.15 and is observed to be about 15\% higher with respect to that of Silber {\it et al.}\cite{silber17}. Interestingly the ratios are found to remain constant with respect to pressure for both $Fe$ and $Ni$  along the melting boundary. The low thermal conductivity observed in the MAC experiments may result from high resistivity due to diffusion of W or Re from the thermocouple during melting of the sample\cite{silber17,silber18}  as can be seen from Fig.6(b) in their study. It was reported that the liquid $Fe$ maintain a local closed-packed hard-sphere structure and it remains invariant along its melting curve\cite{shen04}. Silber {\it et al.}\cite{silber17} discussed about having a P-invariant Fermi surface and constant electron mean free path at the onset of melting in $Ni$. This may be a reasonable explanation for observing the constant value of thermal conductivity of $Fe$ and $Ni$ along its melting curve, because both are transition metals having unfilled 3d cell and similar electronic configuration. In the other study of the resistivity measurements on $Ni$, a decrease in the resistivity during melting was observed\cite{silber17} in a pressure range is from 0 to 9 GPa. Similar to the above phenomena, we have also observed a linear increase in the $\kappa$-values of $Ni$ with pressure while they are plotted in a near isotherm (Fig.6(d)). Similar increase is also observed in case of $Fe$, which is shown in Fig.6(c) consistent with our previous study\cite{sahagf20}.

$Cu$ has filled d cell, shows much steeper melting curve with respect to unfilled d cell containing materials such as $Fe$ and $Ni$\cite{japel05,ross07,errandonea13}. $Fe$ and $Ni$ has similarities in the magnetic states and it has been observed that alloying of $Ni$ with $Fe$ at 5.5\% remains in the same hcp structure at high pressure and this structure is more stable than pure $Fe$\cite{lin02,tateno12}. Also in ab initio calculations, the seismic properties of $Fe$ and $Ni$ alloy are observed to be almost indistinguishable from those of pure $Fe$\cite{davies15,martorell13}. Due to the above mentioned similar properties, both $Fe$ and $Ni$ exhibit anomalous shallow melting curve\cite{japel05} attributed to their d-electrons by Japel {\it et al}. From the sudden decrease in the $\kappa$-values we have calculated the melting temperatures at high pressures for $Fe$ and $Ni$ since at those temperature points there are no other structural or magnetic transitions. We have compared our measured melting temperature of $Fe$ and $Ni$ in Fig.7(a) and (b), respectively with those with literature values. Melting temperatures of $Fe$ are found to have a very good agreement with the laser heated diamond anvil cell data\cite{liu75,boehler93} and with the later work by Strong {\it et al.}\cite{strong73}. This melting curve is observed to deviate by a very little amount with respect to the previous work by Strong\cite{strong59} and the multi-anvil experiments by Silber {\it et al.}\cite{silber18}.  For $nickel$, melting curve is found to be in very good agreement with all other laser heated and multi-anvil cell experiments\cite{strongbuddy59,japel05,errandonea13,silber17} except that reported by Lazor {\it et al.}\cite{lazor93}. Slight difference in the melting curve with Lazor {\it et al.} may be due to the different fitting procedures for the temperature measurements\cite{lazor93}. We fit the Planck's radiation function in the wavelength range 650 nm - 900 nm taking constant emissivity with respect to wavelength, while Lazor {\it et al.} fitted by a least squares method to Wien's approximation of Planck's radiation function.  It is predicted that the energy of the liquid state at the onset of melting occurs due to the partially filled d-shells and it leads to a loss of d-band structural periodicity as compared to filled d-band metals\cite{japel05,ross07}. It results in a anomalies in compressibility and internal pressure in liquid $Fe$ and $Ni$\cite{steinemann88}. Hence all the above facts play important role in the observed shallow melting curve with pressure for partially filled d-shell metals.

The liquid outer core of the Mercury generates a weak dynamo like Earth\cite{dumberry15} and it consists of molten $Fe$. The pressure and the temperature ranges of the Mercury core mantle boundary is considered to be in the range 5 to 8 GPa and 1850-2200 K. Our experimental pressure and temperature ranges for $Fe$ belong to same range and we find that $Fe$ has a thermal conductivity value of 60-70$\pm$20 $Wm^{-1}K^{-1}$ at their melting and it remains constant over the pressure range we studied. Apart from that we find thermal conductivity of $Ni$ at melting in the range 65-70$\pm$20 $Wm^{-1}K^{-1}$ and it remains constant over the pressure range 4-22 GPa. Since the interior of the planets may have alloys of $Fe$ and $Ni$, the  constant behaviour of thermal conductivity values during melting put an important constrain over the heat conduction of the planetary interiors, specially planets like Mercury and Mars. Though more experiments and theoretical calculations are needed on the Fe, and its alloys. 

\section{conclusion}
We have carried out the direct measurements of the thermal conductivity ($\kappa$) of $Fe$ and $Ni$ along its melting curve. Near-isothermal $\kappa$'s are observed to increase with pressure in both the metals due to the increase of density of the pressed metals. A sudden decrease of $\kappa$ with temperature is observed for both the metals, and these temperatures for different pressures are attributed to the melting of the respective metals. Melting curve for both the metals are found to agrees very well with the other multi-anvil and laser heated experiments. Thermal conductivity values for both the metals at their melt are observed to remain constant with pressure in consistent with the other multi-anvil measurements. Constant values of $\kappa$ in these metals during melting at different pressures can be attributed due to the constant Fermi surface and an invariant
electron mean free path in melt for both the metals. Melting gives rise to loss of long range order and maintains a local closed-packed hard-sphere structure, which remains invariant along its melting curve resulting in an independent movement of atomic metals. The above phenomena is observed in the unfilled d-band transition metals.

\section*{Acknowledgement(s)}

\section*{Funding}

GDM wishes to thank Ministry of Earth Sciences, Government of India for financial support under the project grant no. MoES/16/25/10-RDEAS. PS wishes to thank DST, INSPIRE program by Department of Science and Technology, Government of India for financial support.

\section*{Notes on contributor(s)}

All authors has equal contribution. Both authors reviewed the manuscript.
\section{Additional Information}


\newpage
\begin{table}
\caption{\label{tab:table1} The parameter values used in COMSOL for determination of the thermal conductivity
of $Fe$ and $Ni$.}

\begin{tabular}{cccc}\\
\hline
 Material & Dimensions & Density & Thermal Conductivity ($\kappa$)\\
&Thickness, and diameter ($\mu$m)& ($Kg.m^{-3}$) & $W m^{-1} K^{-1}$\\
\hline
$Fe$ & 15, and 90-110 & 7620& Variable\\
$Ni$ & 15, and 100-120 & 8485& Variable\\
$NaCl$ & 12, and 110-120 & 2160& 6 \cite{hakanson86}\\
Gasket (Steel) & 40-45, and 10$^6$ & 8050& 20 \cite{kiefer05}\\
\hline
\end{tabular}
\end{table}

\newpage
\begin{table}
\caption{\label{tab:table2} Detailed error analysis in the measurement of the thermal conductivity of $Ni$ plate at a pressure of 17.2 GPa. The radius of hotspot is $r_1$, the ambient thickness of the $Ni$ plate is $h$, mass of the hotspot is definedas $m$, specific heat of $Ni$ is $c$. $T_1$ is the temperature of the hotspot, $T_2$ is the temperature at a distance $r_2$ from the center of hotspot, $Q$ is the absorbed power at hotspot measured using Eqn.4, and $k$ is thermal conductivity measured using Eqn.3. The error in $r_1$ is assigned from the difference of the half of beam waist of the incident infra-red laser and the radius of the hotspot. The error in determination of the thickness ($h$) of the compressed $Ni$-plate is calculated from the several measurements of thickness before loading the sample. The error in mass ($m$) of the hotspot is measured from the error measurements of density and volume of the hotspot assuming the quasi-hydrostatic condition. The error in $C_p$ is assigned from the literature\cite{nickelc}. The error in $T_1$ and $T_2$ is already explained above and is taken to be 50 K. The error in $Q$ is assigned from the propagation of the errors. The error in $r_2$ is assigned from the resolution of the motion of spectrometer pinhole. The error in $k$ is assigned from all the propagated errors. A total error is estimated to be 30$\%$ in the thermal conductivity values.}

\begin{tabular}{cccccccccc}\\
\hline
 &$r_1$& $h$& $m$& $C_p$& $T_1$& $T_2$& $Q$& $r_2$ &$k$\\
&$\mu$m& $\mu$m& $Kg$& $JKg^{-1}K^{-1}$& $K$&$K$& $Watt$& $\mu$m& $W m^{-1} K^{-1}$\\
\hline
Measured &9& 15& 6.2$\times10^{-11}$& 420& 1740& 1467& 1.9& 70& 103\\

value & &  &   &   &    &    &    &   & \\
Error &$\pm$2& $\pm$0.2& $\pm$1.4$\times10^{-11}$& $\pm$5& $\pm$50& $\pm$50& $\pm$0.5& $\pm$1& $\pm$31\\
\hline
\end{tabular}
\end{table}

\newpage

\begin{figure}
\includegraphics[scale=1]{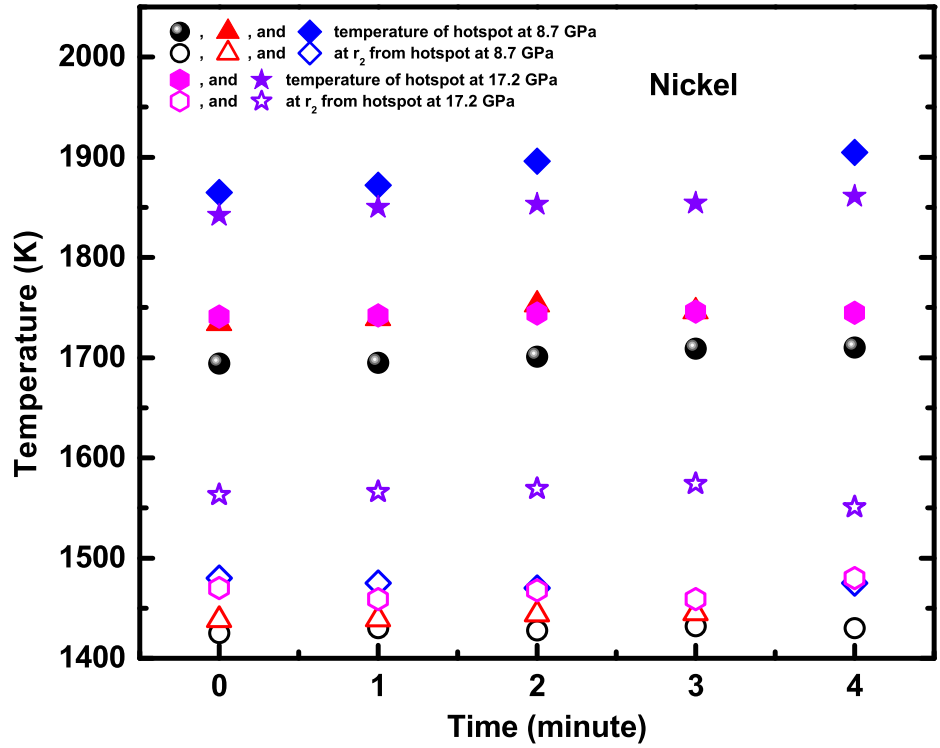}
\caption{ Time dependent temperatures at the hotspot (filled symbols) and at a distance $r_2$ (70 $\mu$m) from the hotspot (open symbols) on a $Ni$ plate at two pressure points while heated at different temperatures.}
\end{figure}

\newpage
\begin{figure}
\includegraphics[scale=1]{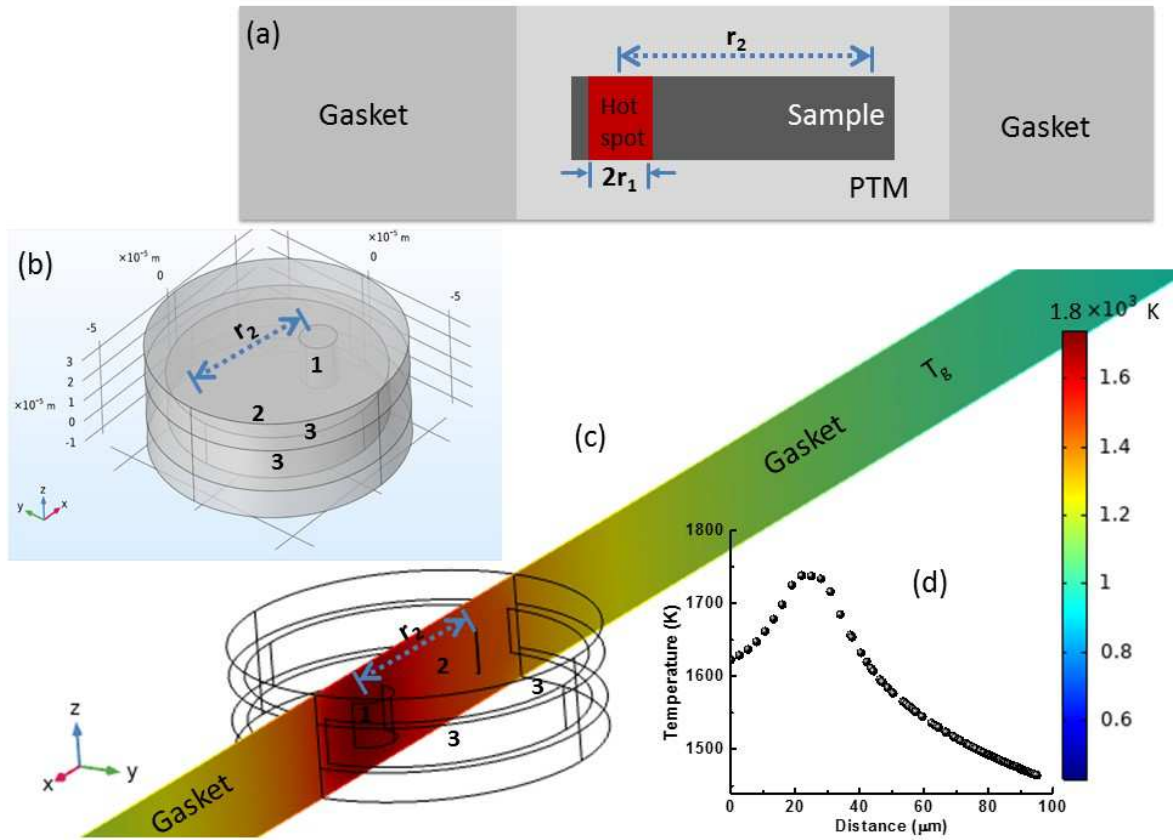}
\caption{ (a) Schematic cross sectional view of the sample chamber with gasket during heating of $Ni$ plate at one of its edge. (b) Schematic drawing of the sample chamber geometry inside the gasket hole of the LHDAC having 300 micron culet in a COMSOL software. Number 1 represents the hotspot at the one of edge of the $Ni$ plate, 2 represents the compressed $Ni$ plate, and 3 represents pressure transmitting medium (PTM). (c) Cross sectional view of computed temperature distribution in the sample chamber and gasket material while the hotspot temperature is 1740 K and $Ni$ thermal conductivity is 103 $Wm^{-1}K^{-1}$. (d) Computed line profile of the temperature on the $Ni$ plate for the above condition.}
\end{figure}

\newpage
\begin{figure}
\includegraphics[scale=1]{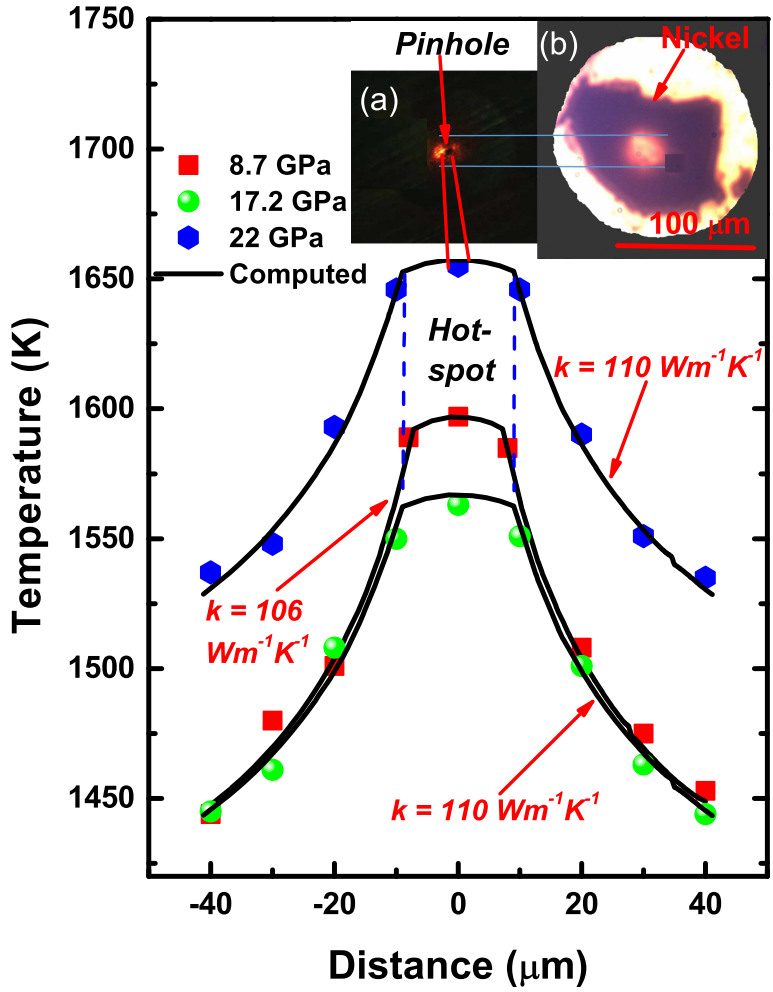}
\caption{ Measured and computed temperature distribution on the $Ni$ plate heated at different pressures. Temperatures were measured by translating the 50 $\mu$m pinhole attached to the spectrometer across the magnified image (magnified by 16 times) of the sample surface. Inset (a) Shows the magnified image of the 50 $\mu$m pinhole while heating the $Ni$ plate. Pinhole captures thermal radiation of 3 $\mu$m of the sample surface and (b) shows the magnified image of the $Ni$ loaded sample chamber at a pressure 17.2 GPa heated at 1563 K under transmitting light. The redish glow in both the inset is the hotspot about diameter 18 $\mu$m.}
\end{figure}

\newpage
\begin{figure}
\includegraphics[scale=1]{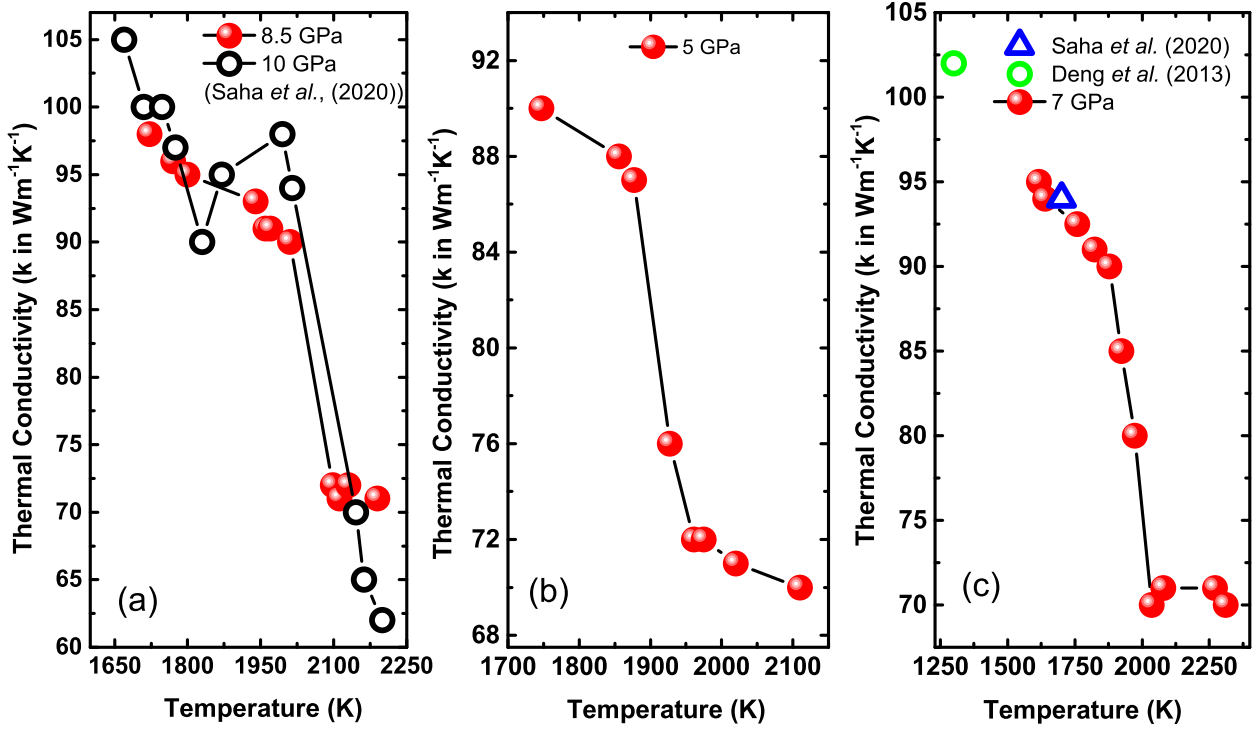}
\caption{ The comparison of temperature dependent thermal conductivity of $Fe$ at different pressures. (a) Represents the data of Saha {\it et al.}\cite{sahagf20} at a pressure 10 GPa. (b), (c), and (d) Represents the data obtainted in this work at 5, 7, and 8.5 GPa respectively. In (c), green open circle and blue open triangle data are from electrical resistivity measurements by Dang {\it et al.}\cite{deng13} and direct measurements using LHDAC by Saha {\it et al.}\cite{sahagf20} around 7 GPa, respectively. At each pressure,  after certain temperature, $\kappa$ shows a sudden drop. These transition temperature values are in well agreement with the melting points of $Fe$ at the respective pressures.\cite{liu75,anderson86}}   
\end{figure}

\newpage
\begin{figure}
\includegraphics[scale=1]{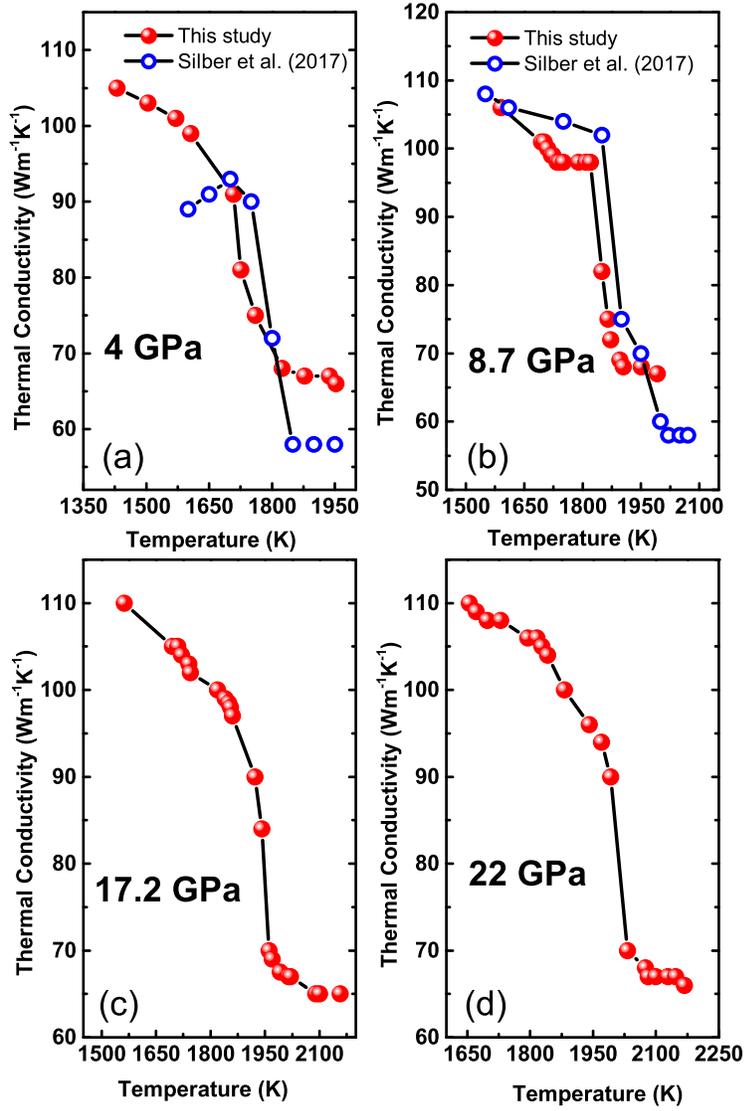}
\caption{ The comparison of temperature dependent thermal conductivity of $Ni$ at different pressures. In (a), and (b) all filled red circles represent our data at 4, and 8.7 GPa, respectively while green open circles represent the data by Silber {\it et al.}\cite{silber17} from electrical resistivity measurements in a MA large volume press at 4 and 9 GPa. In both data in (a), and (b) $\kappa$ show a sharp fall after certain temperature and reported to be melting\cite{silber17}. (c), (d) Represent temperature dependent $\kappa$ of $Ni$ in our study at 17.2 and 22 GPa. Melting induced sharp fall in $\kappa$ is evident in (c), and (d).}
\end{figure}

\newpage
\begin{figure}
\includegraphics[scale=1]{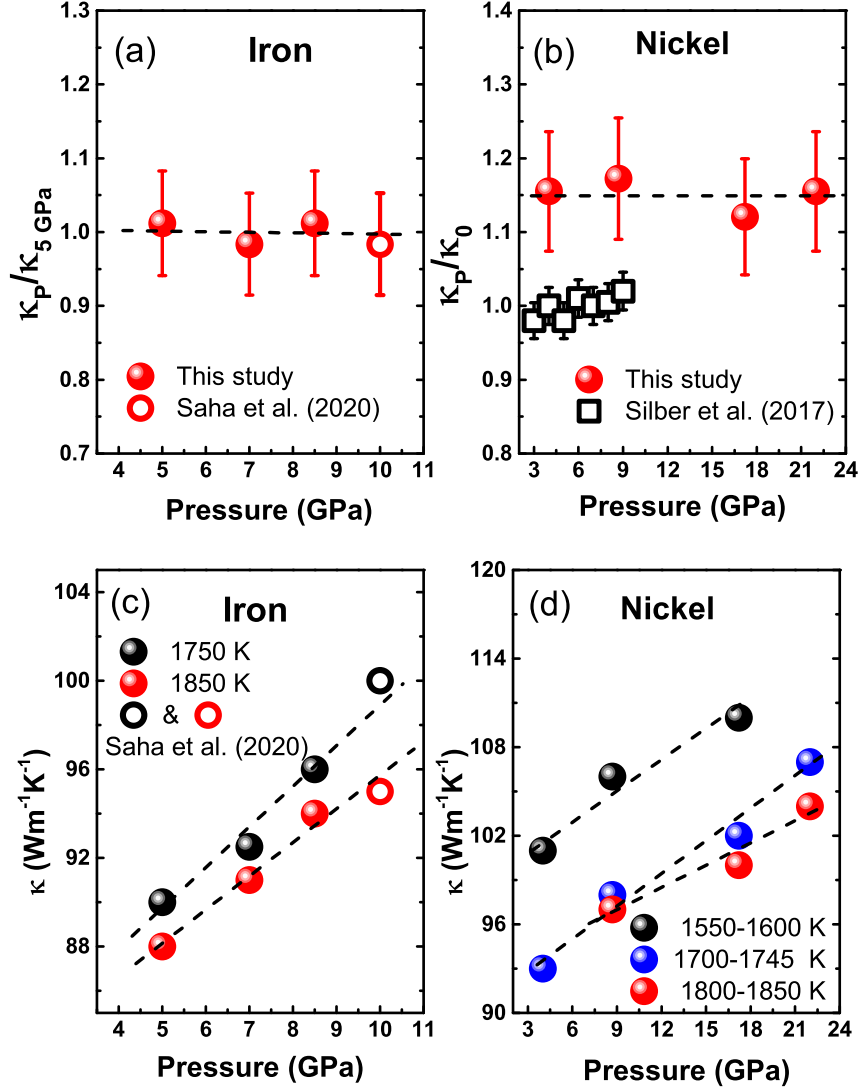}
\caption{ (a) The ratio of pressure dependent $\kappa_P$ to that at 5 GPa $\kappa_5$ of $Fe$ during melting at hotspot shows a constant value. Filled circles represent our data and open circle represent data of Saha {\it et al.}\cite{sahagf20}. (b) Comparison of ratio of pressure dependent $\kappa_P$ to that at ambient pressure $\kappa_0$ of $Ni$ during melting. Filled circles represent our data and open squares represent data of Silber {\it et al.}\cite{silber17}. Both data shows a constant value of these ratio. Errors in our data are assigned from the deviation of $\kappa$ values during melting at the respective pressures. (c), (d) Represent the pressure dependent near isothermal thermal conductivity of $Fe$, and $Ni$, respectively. All the filled symbols represent our data while all the open symbols are the literature values of $\kappa$\cite{sahagf20}.}
\end{figure}

\newpage
\begin{figure}
\includegraphics[scale=1]{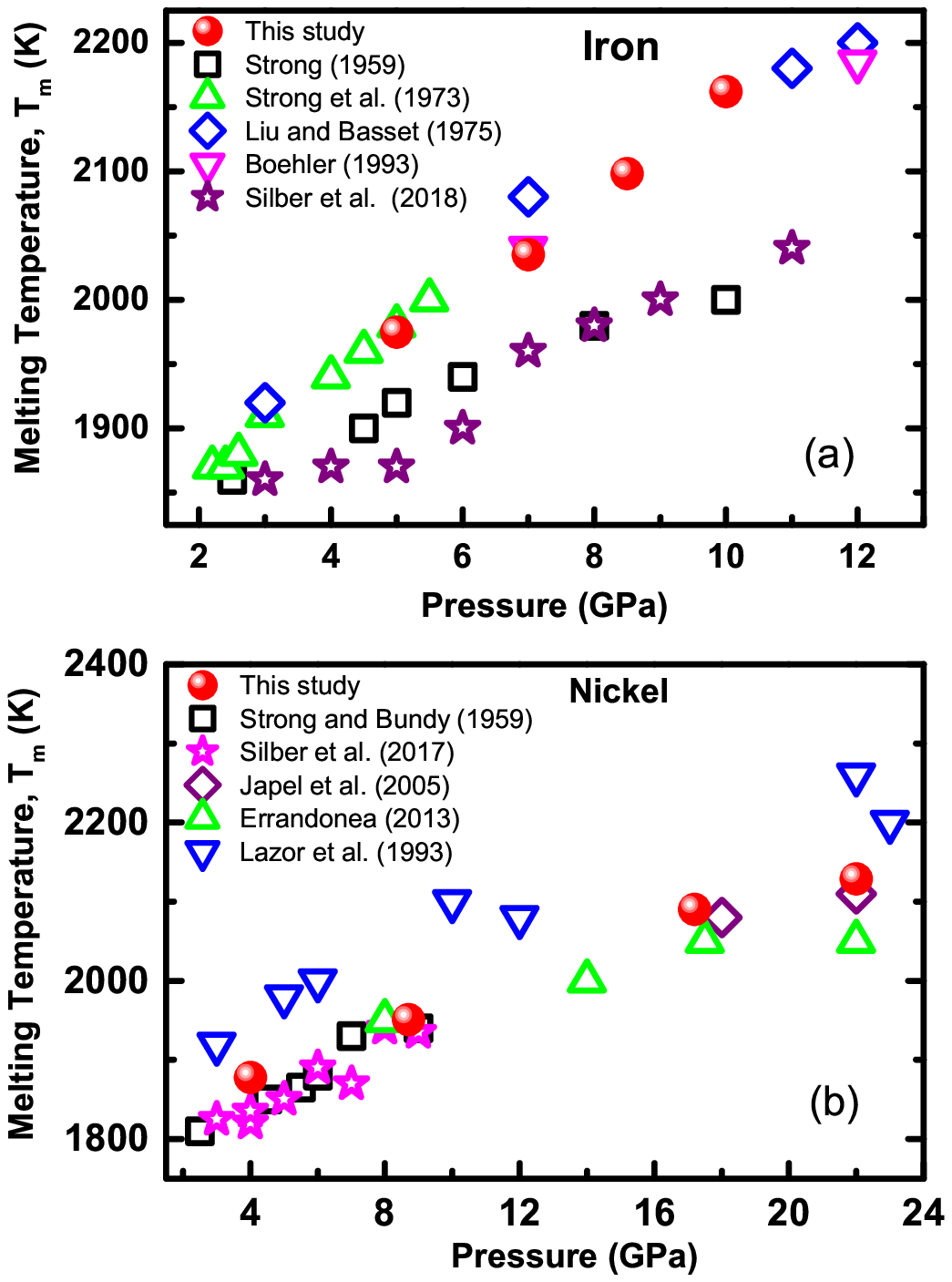}
\caption{ Comparison of the melting temperature with pressure estimated from the observation of sharp fall in the $\kappa$'s values with the other measurements\cite{strong59,strong73,liu75,boehler93,silber18,strongbuddy59,silber17,japel05,errandonea13,lazor93}: (a) $Fe$, (b) $Ni$, respectively. All the filled symbols represent our data while all the open symbols are the reported data\cite{strong59,strong73,liu75,boehler93,silber18,strongbuddy59,silber17,japel05,errandonea13,lazor93} using different techniques.}
\end{figure}

\begin{thebibliography}{}
\bibitem{birch52}
F. Birch, Elasticity and constitution of the Earth's interior. {\it J. Geo-phys. Res.} {\bf57}, 227-286 (1952).  https://doi.org/10.1029/JZ057i002p00227

\bibitem{dziewonski81}
A. M. Dziewonski, and D. L. Anderson, Preliminary reference Earth model. {\it Phys. Earth Planet inter.} {\bf25(4)}, 297-356 (1981). https://doi.org/10.1016/0031-9201(81)90046-7

\bibitem{stevenson81}
D. J. Stevenson, Models of the Earth's Core. {\it Science} {\bf214}, 611-619 (1981). https://doi.org/10.1126/science.214.4521.611


\bibitem{mao90}
H. K. Mao, Y. Wu, L. C. Chen, J. F. Shu, and A. P. Jephcoat, Static compression of iron to 300 GPa and Fe0.8Ni0.2 alloy to 260 GPa: Implications for composition of the core. {\it J. Geophys. Res.} {\bf95}, 21,737-21,742 (1990). https://doi.org/10.1029/JB095iB13p21737


\bibitem{wood93}
B. J. Wood, Carbon in the core. {\it Earth Planet. Sci. Lett.} {\bf 117(3-4)}, 593-607 (1993). https://doi.org/10.1016/0012-821X(93)90105-I

\bibitem{poirier94}
J. P. Poirier, Light elements in the Earth's outer core: A critical review. {\it Phys. Earth Planet. Inter.} {\bf85}, 319-337 (1994). https://doi.org/10.1016/0031-9201(94)90120-1

\bibitem{dubrovinsky00}
L. S. Dubrovinsky, S. K. Saxena, F. Tutti, S. Rekhi, and T. Lebehan, In Situ X-Ray Study of Thermal Expansion and Phase Transition of Iron at Multimegabar Pressure. {\it Phys. Rev. Lett.} {\bf84}, 1720-1723 (2000). https://doi.org/10.1103/PhysRevLett.84.1720

\bibitem{li02} 
J. Li, H. K. Mao, Y. Fei, E. Gregoryanz, M. Eremets, and C. S. Zha, Compression of $Fe_3C$ to 30 GPa at room temperature. {\it Phys. Chem. Miner.} {\bf29(3)}, 166-169 (2002). https://doi.org/10.1007/s00269-001-0224-4


\bibitem{li03}
J. Li, and Y. Fei, Experimental constraints on core composition. in Treatise on Geochemistry (2nd edition), edited by H. D. Holland and K. K. Turekian, pp. 527-557, Elsevier, Oxford (2003).


\bibitem{nakajima09}
Y. Nakajima, E. Takahashi, T. Suzuki, and K. Funakoshi, "Carbon in the core" revisited. {\it Phys. Earth Planet. Inter.} {\bf174}, 202-211 (2009). https://doi.org/10.1016/j.pepi.2008.05.014

\bibitem{sata10}
N. Sata, K. Hirose, G. Shen, Y. Nakajima, Y. Ohishi, and N. Hirao, Compression of $FeSi$, $Fe_3C$, $Fe_{0.95}O$, and $FeS$ under the core pressures and implication for light element in the Earth’s core. {\it J. Geophys. Res.} {\bf115}, B09204 (2010). https://doi:10.1029/2009JB006975


\bibitem{mookherjee11}
M. Mookherjee, Y. Nakajima, G. Steinle-neumann, K. Glazyrin, X. Wu, L. Dubrovinsky, C. McCammon, and A. Chumakov, High‐pressure behavior of iron carbide ($Fe_7C_3$) at inner core conditions. {\it J. Geophys. Res.} {\bf116}, B04201 (2011). doi:10.1029/2010JB007819,


\bibitem{nakajima11}
Y. Nakajima, E. Takahashi, N. Sata, Y. Nishihara, K. Hirose, K. Funakoshi, and Y. Ohishi, Thermoelastic property and high-pressure stability of $Fe_7C_3$: Implication for iron-carbide in the Earth's core. {\it Am. Mineral.} {\bf 96}, 1158-1165 (2011).  https://doi.org/10.2138/am.2011.3703

\bibitem{chen12}
B. Chen, L. Gao, B. Lavina, P. Dera, E. E. Alp, J. Zhao, and J. Li, Magneto-elastic coupling in compressed $Fe_7C_3$ supports
carbon in Earth’s inner core. {\it Geophys. Res. Lett.} {\bf 39}, L18301 (2012). doi:10.1029/2012GL052875

\bibitem{chen14}
B. Chen, Z. Li, D. Zhang, J. Liu, M. Y. Hu, J. Zhao, W. Bi, E. E. Alp,Y. Xiao, P. Chow, and J. Li, Hidden carbon in Earth’s inner core revealed by shear softening in dense $Fe_7C_3$. {\it Proceedings of the National Academy of Sciences} {\bf 111(50)}, 17755-17,758 (2014). https://doi.org/10.1073/pnas.1411154111


\bibitem{prescher15}
C. Prescher, L. Dubrovinky, E. Bykova, I. Kupenko, K. Glazyrin, A. Kantor, C. McCammon, M. Mookherjee,  N. Nakajima, N. Miyajima, R. Sinmyo, V. Cerantola, N. Dubrovinskaia, V. Prakapenka, R. Ruffer, A. Chumakov, and M. Hanfland, High Poisson's ratio of Earth's inner core explained by carbon alloying. {\it Nat. Geosci.} {\bf 8}, 220 (2015). https://doi.org/10.1038/ngeo2370


\bibitem{liu16}
J. Liu, J. Li, and D. Ikuta, Elastic softening in $Fe_7C_3$ with implications for Earth's deep carbon reservoirs. {\it J. Geophys. Res. Solid Earth} {\bf 121}, 1514-1524 (2016). https://doi.org/10.1002/2015JB012701


\bibitem{saha20}
P. Saha, K. Glazyrin, G. D. Mukherjee, Synthesis and Compression study of orthorhombic $Fe_7(C,Si)_3$: A possible constituent of the Earth's core. arXiv preprint arXiv:1905.11030

\bibitem{stacey01} 
F. D. Stacey, and  O. L. Anderson, Electrical and thermal conductivities of $Fe-Ni-Si$ alloy under core conditions. {\it Phys. Earth Planet Inter.} {\bf 124}, 153-162 (2001). https://doi.org/10.1016/S0031-9201(01)00186-8

\bibitem{stacey07} 
F. D. Stacey, and D. E. Loper, A revised estimate of the conductivity of iron alloy at high pressure and implications for the core energy balance. {\it Phys. Earth Planet. Inter.} {\bf 161}, 13-18 (2007). https://doi.org/10.1016/j.pepi.2006.12.001

\bibitem{sha11}  
X. Sha, and R. E. Cohen, First-principles studies of electrical resistivity of iron under pressure.  {\it J. Phys. Condens. Matter.} {\bf 23} 075401 (2011). https://doi.org/10.1088/0953-8984/23/7/075401

\bibitem{alfe12}
D. Alfe, M. Pozzo, and M. P. Desjarlais, Lattice electrical resistivity of magnetic bcc iron from first-principles calculations. {\it Phys. Rev. B.} {\bf 85}, 024102(2012). https://doi.org/10.1103/PhysRevB.85.024102

\bibitem{deKoker12} 
N. de Koker, G. Steinle-Neumann, and V. Vlcek, Electrical resistivity and thermal conductivity of liquid $Fe$ alloys at high P and T, and heat flux in Earth's core.  {\it PNAS}. {\bf 109} 4070-4073 (2012). https://doi.org/10.1073/pnas.1111841109


\bibitem{pozzo12} 
M. Pozzo, C. Davies, D. Gubbins, and D. Alfe, Thermal and electrical conductivity of iron at  Earth's core conditions. {\it Nature}. {\bf 485}, 355-358 (2012). https://doi.org/10.1038/nature11031

\bibitem{pozzo14}
M. Pozzo, C. Davies, D. Gubbins, and D. Alfe, Thermal and electrical conductivity of solid iron and iron–silicon mixtures at Earth's core conditions. {\it Earth Planet. Sci. Lett.} {\bf 393}, 159-164 (2014). https://doi.org/10.1016/j.epsl.2014.02.047

\bibitem{davies15} 
C. Davies, M. Pozzo, D. Gubbins, and D. Alfe, Constraints from material properties on the dynamics and evolution of Earth's core. {\it Nature Geoscience}. {\bf 8}, 678-685 (2015). https://doi.org/10.1038/ngeo2492

\bibitem{xu18}
J. Xu, P. Zhang, K. Haule, J. Minar, S. Wimmer, H. Ebert, R. E. Cohen, Thermal conductivity and electrical resistivity of solid iron at Earth’s core conditions from first principles. {\it Phys. Rev. Lett.} {\bf 121} (9) 096601 (2018). https://doi.org/10.1103/ PhysRevLett.121.096601.

\bibitem{zidane20}
M. Zidane, E. M. Salmani, A. Majumdar, H. Ez-Zahraouy, A. Benyoussef, and R. Ahuja, Electrical and thermal transport properties of $Fe–Ni$ based ternary alloys in the earth's inner core: An ab initio study. {\it Phys. Earth Planet. Inter.} {\bf 301}, 106465 (2020). https://doi.org/10.1016/j.pepi.2020.106465


\bibitem{gomi13} 
H. Gomi, K. Ohta, K. Hirose, S. Labrosse, R. Caracas, M. J. Varstraete, and J. W. Hernlund, The high conductivity of iron and thermal evolution of the Earth's core. {\it Phys. Earth Planet. Inter.} {\bf 224}, 88-103 (2013). http://dx.doi.org/10.1016/j.pepi.2013.07.010

\bibitem{ohta16} 
K. Ohta, Y. Kuwayama, K. Hirose, K. Shimizu, and Y. Ohishi, Experimental determination of  the electrical resistivity of iron at Earth's core conditions. {\it Nature}. {\it 534}, 95-98 (2016). https://doi.org/10.1038/nature17957


\bibitem{konopkova11} 
Z. Konpokova, P. Lazor, A. F. Goncharov, and V. V. Struzhkin, Thermal conductivity of hcp iron  at high pressure and temperature. {\it High Press. Res.} {\bf 31}, 228-236 (2011). https://doi.org/10.1080/08957959.2010.545059

\bibitem{konopkova16} 
Z. Konopkova, R. S. McWilliams, N. Gomez-Perez, and A. F. Goncharov, Direct measurement of thermal conductivity in solid iron at planetary core conditions. {\it Nature}. {\it 534}, 99-101 (2016). https://doi.org/10.1038/nature18009

\bibitem{sahagf20}
P. Saha, A. Mazumder, and G. D. Mukherjee. Thermal conductivity of dense hcp $iron$: Direct measurements using laser
heated diamond anvil cell. {\it Geoscience Frontiers.} {\bf 11}, 1755-1761 (2020). https://doi.org/10.1016/j.gsf.2019.12.010


\bibitem{deng13} 
L. Deng, C. Seagle, Y. Fei, and A. Shahar, High pressure and temperature electrical resistivity of iron and implications for planetary cores. {\it Geophys. Res. Lett.} {\bf 40}, 33-37 (2013). https://doi.org/10.1029/2012GL054347


\bibitem{secco17}
R. A. Secco, Thermal conductivity and Seebeck coefficient of $Fe$ and $Fe-Si$ alloys: Implications for variable Lorenz number. {\it Earth Planet. Sci. Lett.} {\bf 264}, 23-34 (2017). https://doi.org/10.1016/j.pepi.2017.01.005


\bibitem{silber17}
R. E. Silber, R. A. Secco, and W. Yong, Constant electrical resistivity of $Ni$ along the melting boundary up to 9 GPa. {\it J. Geo. Res.: Solid Earth.}  {\bf 122}, 5064-5081 (2017).  https://doi.org/10.1002/2017JB014259


\bibitem{silber18}
R. E. Silber, R. A. Secco, W. Yong, and J. A. H. Littleton, Electrical resistivity of liquid $Fe$ to 12 GPa: Implications for heat flow in cores of terrestrial bodies. {\it Sci. Rep.} {\bf 8}, 10758 (2018). https://doi.org/10.1038/s41598-018-28921-w


\bibitem{ezenwa19}
I. C. Ezenwa, and R. A. Secco, $Fe$ melting transition: Electrical resistivity, thermal conductivity, and heat flow at the inner core boundaries of mercury and ganymede. {\it Crystals.} {\bf 9}, 359 (2019). https://doi.org/10.3390/cryst9070359


\bibitem{basu20}
A. Basu, M. R.Field, D. G. McCulloch, and R. Boehler, New measurement of melting and thermal conductivity of iron close to outer core conditions. {\it Geoscience Frontiers.} {\bf 11}, 565-568 (2020). https://doi.org/10.1016/j.gsf.2019.06.007

\bibitem{comsol1}
COMSOL Multiphysics reference manual, version 5., COMSOL, Inc, {\it www.comsol.com.}

\bibitem{mao86} 
H. K. Mao, J. Xu, and P. M. Bell, Calibration of the Ruby pressure Gauge to 800 kbar Under  Quasi-Hydrostatic Conditions. {\it J. Geophys. Res.} {\bf 91}, 4673-4676 (1986). https://doi.org/10.1029/JB091iB05p04673

\bibitem{planck01} 
M. Planck, On the Law of the Energy Distribution in the Normal Spectrum. {\it Ann. Physik}. {\bf 4}, 553 (1901).


\bibitem{boehler90} 
R. Boehler, N. V. Bargen, and A. Chopelas, Melting, Thermal Expansion, and Phase Transitions of Iron at High Pressures. {\it J. Geophys. Res.} {\bf 95}, 21,731-21,736 (1990). https://doi.org/10.1029/JB095iB13p21731


\bibitem{mukherjee07} 
G. D. Mukherjee, and R. Boehler, High-pressure Melting Curve of Nitrogen and the Liquid-Liquid  Phase Transition. {\it Phys. Rev. Lett.} {\bf 99}, 225701 (2007). https://doi.org/10.1103/PhysRevLett.99.225701

\bibitem{sahaijp20}
P. Saha, and G. D. Mukherjee, Temperature measurement in double-sided laser-heated diamond anvil cell and reaction of carbon. {\it Indian J Phys} (2020), https://doi.org/10.1007/s12648-020-01699-2


\bibitem{gubbins03} 
D. Gubbins, D. Alfe, G. Masters, G. D. Price, and M. J. Gillan, Can the Earth's dynamo run  on heat alone? {\it Geophys. J. Int.} {\bf 155}, 609-622 (2003). https://doi.org/10.1046/j.1365-246X.2003.02064.x

\bibitem{hirose13} 
K. Hirose, S. Labrosse, and J. Hernlund, Composition and State of the Core. {\it Annu. Rev. Earth. Planet. Sci.} {\bf 41}, 657-691 (2013). https://doi.org/10.1146/annurev-earth-050212-124007

\bibitem{nickelc}
A. Cezairliyan, and A. P. Miiller, Heat Capacity and Electrical Resistivity of Nickel in the Range 1300-1700 K Measured with a
Pulse Heating Technique. {\it International Journal of Thermophysics.} {\bf 4}, 4 (1983). doi:10.1007/BF01178788


\bibitem{hakanson86}
B. Hakansson, P. Andersson, Thermal Conductivity and heat capacity of solid NaCl and NaI under pressure. {\it J. Phys. Chem. Solids.} {\bf 47 (4)}, 355–362 (1986). https://doi.org/10.1016/0022-3697(86)90025-9.


\bibitem{kiefer05}
B. Kiefer, T. S. Duffy, Finite element simulations of the laser-heated diamond-anvil cell. {\it J. Appl. Phys.} {\bf 97 (11)}, 114902 (2005). https://doi.org/10.1063/1.1906292.


\bibitem{liu75} 
L. G. Liu, and  W. A. Bassett, The Melting of Iron up to 200 kbar. {\it J. Geophys. Res.} {\bf 80}, 3777-3782 (1975). https://doi.org/10.1029/JB080i026p03777

\bibitem{anderson86} 
O. L. Anderson, Properties of iron at the Earth's core conditions.  {\it Geophys. J. R. astr. Soc.} {\bf 84}, 561-579 (1986). https://doi.org/10.1111/j.1365-246X.1986.tb04371.x

\bibitem{ho72} 
C. Y. Ho, R. W. Powell, and P. E. Liley, Thermal conductivity of the elements. {\it J. Phys. Chem.  Ref.  Data}. {\bf 1}, 279-422 (1972). https://doi.org/10.1063/1.3253100

\bibitem{nishi03} 
T. Nishi, H. Shibata, Y. Ohta, and Y. Waseda, Thermal Conductivity of Molten Iron, Cobalt, and Nickel by Laser Flash Method. {\it Metall. Mater. Trans. A}. {\bf 34}, 2801-2807 (2003). https://doi.org/10.1007/s11661-003-0181-2


\bibitem{chuandchi81}
T. K. Chu, and T. C. Chi, Properties of selected Ferrous Alloying Elements, lll-1, 269 pp., McGraw-Hill, Washington.



\bibitem{strong59}
H. M. Strong The Experimental Fusion Curve of Iron to 96,000 Atmospheres. {\it Journal of Geophysical Research.} {\bf 64}, 653–659(1959). https://doi.org/10.1029/JZ064i006p00653.


\bibitem{strong73}
H. M. Strong, R. E. Tuft,and R. E. Hanneman The iron fusion curve and $\gamma$-$\delta$-l triple point. {\it Metallurgical Transactions.} {\bf 4}, 2657–2661 (1973). https://doi.org/10.1007/BF02644272.

\bibitem{boehler93} 
R. Boehler, Temperatures in the Earth's core from melting-point measurements of iron at high  static pressures. {\it Nature}. {\bf 363}, 534-536 (1993). https://doi.org/10.1038/363534a0

\bibitem{strongbuddy59}
H. M. Strong, and F. P. Bundy, Fusion curves of four group VIII metals to 100 000 atmospheres. {\it Phys. Rev.} {\bf 115(2)}, 278–284 (1959). doi:10.1103/PhysRev.115.278.


\bibitem{japel05}
S. Japel, B. Schwager, R. Boehler, and M. Ross, Melting of copper and nickel at high pressure: The role of d electrons. {\it Phys. Rev. Lett.} {\bf 95(16)}, 167801 (2005). doi:10.1103/PhysRevLett.95.167801.

\bibitem{errandonea13}
D. Errandonea, High-pressure melting curves of the transition metals Cu, Ni, Pd, and Pt. {\it Phys. Rev. B.} {\bf 87(5)}, 054108 (2013). doi:10.1103/PhysRevB.87.054108.



\bibitem{lazor93}
P. Lazor, G. Shen, and S. K. Saxena, Laser-heated diamond anvil cell experiments at high pressure: Melting curve of nickel up to
700 kbar. {\it Phys. Chem. Miner.} {\bf 20(2)}, 86–90 (1993). doi:10.1007/BF00207200.

\bibitem{shen04}
G. Shen, V. B. Prakapenka, M. L. Rivers, and S. R. Sutton, Structure of Liquid Iron at Pressures up to 58 GPa. {\it Phys. Rev. Lett.} {\bf 92}, 185701 (2004). doi:10.1103/PhysRevLett.92.185701.


\bibitem{lin02}
J.-F. Lin, D. L. Heinz, A. J. Campbell, J. M. Devine, W. L. Mao, and G. Shen, Iron-nickel alloy in the Earth’s core. {\it Geophys. Res. Lett.} {\bf 29(10)}, 109-101-109-103 (2002). doi:10.1029/2002GL015089.

\bibitem{tateno12}
S. Tateno, K. Hirose, T. Komabayashi, H. Ozawa, and Y. Ohishi, The structure of Fe-Ni alloy in Earth’s inner core. {\it Geophys. Res. Lett.} {\bf 39}, L12305 (2012). doi:10.1029/2012GL052103.

\bibitem{martorell13}
B. Martorell, J. Brodholt, I. G. Wood, and L. Vocadlo, The effect of nickel on the properties of $Fe$ at the conditions of Earth’s inner core: Ab initio calculations of seismic wave velocities of Fe–Ni alloys. {\it Earth Planet. Sci. Lett.} {\bf 365}, 143–151 (2013). doi:10.1016/j.epsl.2013.01.007.

\bibitem{ross07}
M. Ross, R. Boehler, and D. Errandonea, Melting of transition metals at high pressure and the influence of liquid frustration: The late metals Cu, Ni, and Fe. {\it Phys. Rev. B} {\bf 76(18)}, 184117 (2007). doi:10.1103/PhysRevB.76.184117.


\bibitem{steinemann88}
S. Steinemann, and N. Keita, Compressibility and internal pressure anomalies of liquid 3d transition metals. {\it Helv. Phys. Acta} {\bf 61(4)}, 557–565 (1988). doi:10.5169/seals-115960.

\bibitem{dumberry15}
M. Dumberry, and A. Rivoldini, Mercury’s inner core size and core-crystallization regime. {\it Icarus} {\bf 248}, 254–268 (2015). doi.org/10.1016/j.icarus.2014.10.038.


\end{thebibliography}
\end{document}